# A STUDY OF PRE-VALIDATION

By Holger Höfling[1] and Robert Tibshirani[2]

*Stanford University*

Given a predictor of outcome derived from a high-dimensional dataset, pre-validation is a useful technique for comparing it to competing predictors on the same dataset. For microarray data, it allows one to compare a newly derived predictor for disease outcome to standard clinical predictors on the same dataset. We study pre-validation analytically to determine if the inferences drawn from it are valid. We show that while pre-validation generally works well, the straightforward "one degree of freedom" analytical test from pre-validation can be biased and we propose a permutation test to remedy this problem. In simulation studies, we show that the permutation test has the nominal level and achieves roughly the same power as the analytical test.

**1. Introduction.** Suppose that we have a prediction rule derived on a high-dimensional dataset. It is often of interest to compare the new prediction rule to competing rules in order to determine if the new rule provides any additional benefit. For example, the new prediction rule might be based on microarray expression values, while the competing predictors are clinical, nongenomic measurements. Doing the comparison between the new and competing rules on the same dataset (the "re-use" method) would favor the new rule as it was derived on this same dataset. Another approach would be to split the data into separate training and test datasets, build the predictor on the training set and then fit it along with competing predictors on the test set [see Chang et al. (2005) for an example]. However, with limited data, this may severely reduce the accuracy of the new prediction rule and/or the test set may be too small to have adequate power for the comparison.

Pre-validation (PV) [see Tibshirani and Efron (2002)] offers another approach to the problem of comparing a newly derived prediction rule to other

Received July 2007; revised November 2007.
[1]Supported by an Albion Walter Hewlett Stanford Graduate Fellowship.
[2]Supported in part by NSF Grant DMS-99-71405 and NIH Contract N01-HV-28183.

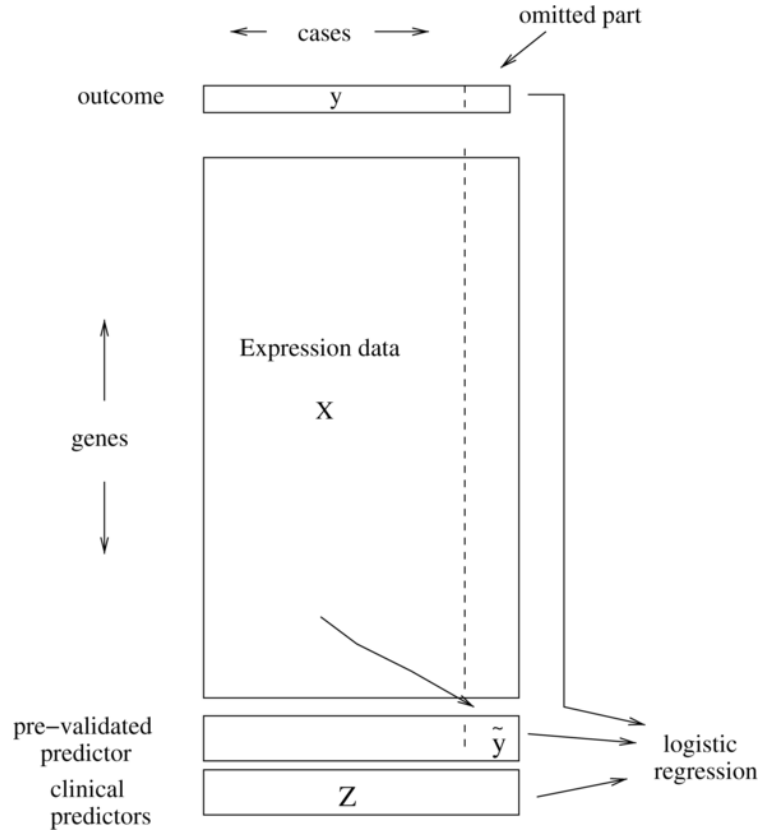

Fig. 1. *A schematic of the pre-validation process. The cases are divided up into (say)* 10 *equal-sized groups. Leaving out one of the groups, a prediction rule is derived from the data of the remaining 9 groups. This prediction rule is then applied to the left out group, giving the pre-validated predictor $\tilde{y}$ for the cases in the left out group. Repeating this process for every group yields the pre-validated predictor $\tilde{y}$ for all cases. Finally, $\tilde{y}$ is included in a logistic regression model together with the clinical predictors to assess its relative strength in predicting the outcome.*

predictors on the same dataset the new rule was derived on. Pre-validation is similar to cross-validation, but instead of directly estimating the prediction error, it constructs a "fairer" version of the predictions on the data. It uses a process similar to cross-validation to construct predictions for each sample, using training features for the other observations. Thus, the result of pre-validation is not an estimate of error (as in cross-validation), but rather a set of pre-validated predictions, one for each sample. These predictions do not have the inherent bias associated with the re-use method. Before going into more details, we explain how pre-validation works on an example (see also Figure 1).

We have microarray data for $n$ patients with breast cancer. On each array, measurements on $p$ genes were taken. Also available are several nonmicroarray based predictors, which are commonly used in clinical practice (e.g., age, tumor size ...) to predict if the patient's prognosis is poor or good. We want to use the microarray data in order to predict the prognosis of a patient. In PV, the $n$ patients are divided into $K$-folds. Leaving out one fold, a prediction rule using the microarray data for the remaining $K-1$ folds is fit (the internal model). Using this rule, the cancer types for the patients in the left out fold are predicted. This way, the data of the left out fold is not used in building the rule and therefore no overfitting occurs. Repeating this procedure for every fold yields a vector of predictions, which we call *pre-validated*. The predicted response for a given patient derives from that patient's covariates through a prediction rule based on independent data. The pre-validated predictor can now be compared to the other nonmicroarray-based predictions using a logistic regression model (the external model). If the coefficient of the pre-validated predictor in the logistic regression model is significant, we conclude that the new microarray-based prediction rule has an independent contribution over the existing rules. The effect of PV is to remove much of the bias that arises from using the same data to build the new prediction rule and compare it to the already established ones.

Pre-validation constructs a fairer version of our predictor that can be used on the same dataset and will act like a predictor applied to a new external dataset. That is, a single pre-validated predictor should behave as if its prediction rule was derived on an independent dataset and therefore act just like a regular predictor in a regression model. In particular, standard tests for significance of the pre-validated predictor should work. In this article we will show that pre-validation is only partially successful: while the coefficient estimate for the pre-validated predictor is generally good, the standard analytical tests (e.g., $t$-test) can be biased, with a level differing from the target level. In this paper we propose a permutation test to solve this problem.

The focus of this paper is the statistical test of significance of the new prediction rule. However, as pointed out in Pepe et al. (2004) and Ware (2006), statistical significance does not necessarily imply scientific or clinical significance. For example, a predictor that improves the misclassification rate of a model from 55% to 60% might be statistical significant, but the overall model might be too inaccurate to use in practice. On the other hand, an improved misclassification rate may appear relevant, however, it may be due to chance and statistically insignificant. Therefore, it is important to establish statistical significance as well as practical relevance. With respect to statistical significance, we can use tests on the pre-validated predictor which should be unbiased to give an accurate answer. In order to establish practical relevance measures like prediction error, true positive fraction (TPF) and false positive fraction (FPF) should be examined. For pre-validation,

Tibshirani and Efron (2002) suggest a method to estimate the improvement in prediction error when using the newly developed prediction method. They show that their method works well and removes much of the bias of the reuse method. Their methodology can easily be extended to other measures such as TPF and FPF. In this article we establish that standard analytical tests in regression models for a pre-validated predictor are biased. This bias may lead researchers to conclude that a new prediction rule is an improvement over other predictors even if the new rule would not have been significant with an unbiased test. We propose a permutation test to remedy this problem.

In Section 3 PV is applied to two different prediction methods on a microarray dataset of breast cancer patients in order to illustrate how it is used in practice. In Section 4 we will establish the bias of the standard test analytically in the simple setting with a linear internal and a linear external model. Section 5 outlines the models that are used in the simulations, the amount of bias of the analytical test in these models and the permutation test. Section 6 presents the results of the simulations.

**2. Pre-validation.** As mentioned above, deriving a prediction rule and comparing it to other rules on the same dataset can lead to a bias in favor of the new rule due to overfitting. This bias can be very large and an example of this effect will be shown later in Section 3.

One way to avoid overfitting is to use separate training and test datasets as in Chang et al. (2005). However, as this is not a very efficient use of the data, we can extend this approach in a straightforward fashion by cross-validation, which is just $K$ applications of the training/test dataset approach. This procedure would then work as follows:

1. Divide the data in $K$ separate groups.
2. Leave out one group and derive the prediction rule over the remaining $K-1$ groups.
3. Using the new prediction rule, predict the outcome for the left out group.
4. Compare the strength of the prediction to the already existing predictors for the outcome (e.g., in a linear or logistic regression model, depending on the type of outcome) only in the left out group. Test if the new predictor is significant.
5. Repeat steps 2–4 for every group and average the results.

However, depending on the choice of $K$, there are tradeoffs. If $K$ is small, say 2 or 3, the prediction rule is derived on a smaller set of data, thus possibly losing accuracy. In situations as with microarray data, where the number of observations is usually small compared to the amount of available data, the reduction of prediction strength due to the lower number of observations can be substantial. On the other hand, if $K$ is, say, 4 or larger, the comparison to the already existing prediction rules has to be done on a very small number of observations. If there are 5 (say) other predictors and a total of 50 observations, then with $K = 5$, the comparison of the new rule to the 5 old ones would have to be done using only 10 observations—it is very unlikely to find significant effects under these circumstances.

Pre-validation (see Figure 1) changes this procedure to avoid the formentioned problems:

1. Divide the data in $K$ separate groups.
2. Leave out one group and derive the prediction rule over the remaining $K - 1$ groups.
3. Using the new prediction rule, predict the outcome for the left out group.
4. Repeat steps 2 and 3 for each group. Collect the predictions into a vector such that one prediction exists for every observation in every group (we call this predictor "pre-validated").
5. Compare the strength of the prediction to the already existing predictors for the outcome (e.g., in a linear or logistic regression model, depending on the type of outcome) using all observations and the predictor derived above. Test if the new predictor is significant.

The main difference of PV to the CV method above is that instead of evaluating the performance on every test set separately and then averaging the results, PV collects a vector with pre-validated predictions for every sample. The comparison to competing predictors is done afterward.

Within PV, any prediction rule can be used, even if the rule itself estimates its parameters by cross-validation. An example of this can be seen in Section 3.2. With respect to the number of folds used in PV, $K$ is usually chosen to be 5 or 10. Leave-one-out PV ($K = n$) leads to high variance in estimates and lower values would decrease the size of the training set too much, as already discussed above. However, as in PV, the predictions for all observations are collected before the comparison to the existing predictors, a high value of $K$ does not compromise the power of this comparison.

When comparing the pre-validated predictor to the existing predictors, usually a linear or logistic regression model is fitted (depending on the outcome). The new prediction rule is judged to make a significant improvement over the old rules if the coefficient of the pre-validated predictor is significantly different from 0. As the new rule predicts the outcome, significant values for the coefficient would be positive. Therefore, instead of a 2-sided test of $\beta_{PV} = 0$ vs. $\beta_{PV} \neq 0$, we can get more power by doing a one-sided test $\beta_{PV} = 0$ vs. $\beta_{PV} > 0$. For this, usually the standard analytical tests for the model (i.e., $t$-statistic or z-score) are used. However, as will be seen in Sections 4 and 6, this analytical test is biased in many situations. We propose to use a permutation test instead, which is explained in detail in Section 5.3 and the performance of which is studied in Section 6. In the next section we want to illustrate how PV works in practice with two examples.

**3. Analysis of breast cancer data.** Here we apply PV to the dataset in van't Veer et al. (2002) using the permutation test and compare it to the analytical results. The data consists of microarray measurements on 4918 genes over 78 patients with breast cancer. Forty-four of these belong to the good prognosis group (survival of more than 5 years), 34 have a poor prognosis. Apart from the microarray data, a number of other clinical predictors exist:

- Tumor grade (good: 1, 2; poor: 3)
- Estrogen receptor (ER) status (good: ≤10; poor: >10)
- Progestron receptor (PR) status (good: ≤10; poor: >10)
- Tumor size (mm) (good: ≤20; poor: >20)
- Patient age (yrs) (good: ≤40; poor: >40)
- Angioinvasion (good: 0; poor: 1)

In order to predict the prognosis of a patient, we try two models. The first has been proposed by van't Veer et al. (2002), the second is a $L_1$ penalized logistic regression model.

3.1. *Van't Veer et al.* (2002) *model.* Based on the microarray data, van't Veer et al. (2002) constructed a predictor for the cancer prognosis in the following way:

1. Select the 70 genes that have the highest correlation with the 78 class labels.
2. Find the centroid vector of the good prognosis group.
3. Compute the correlation of each case with the centroid of the good prognosis group. Find the cutoff such that only 3 cases in the poor prognosis group are misclassified.
4. Classify any new case as good prognosis if their correlation with the centroid is larger than the cutoff.

The predictor from this model is, like the clinical predictors, an indicator variable. Using a continuous response (e.g., probability of being in the poor prognosis group) would be possible, however, we use indicator variables as it is a better match to the clinical predictors, which are also indicators. Using other models is also an option which we explore with the next example. However, as we only want to illustrate how PV works at this point, we do not investigate if there are any other models that possibly give better performance on this dataset. In fact, as we can see in Table 1, the model proposed by van't Veer et al. (2002) performs better than the penalized logistic regression shown below.

An important part of the prediction method is the selection of the top 70 genes. In $K$-fold PV, this selection of the top genes is being repeated separately on each of the $K$ training sets consisting of $(K-1)$ folds as the first step of finding the prediction rule. This way, the top genes used in the prediction are not necessarily the same across the $K$ folds. Zhu, Ambroise and McLachlan (2006) have shown that this reevaluation is important as selecting the top genes in a pre-processing step and keeping them fixed across the folds can lead to biased results.

One option to judge the performance of the new prediction rule is to evaluate its univariate prediction error using CV (see Table 1). However, as we cannot assess this way if two predictors complement each other and therefore improve performance if used together or not, we instead do a multivariate comparison using a logistic regression with the pre-validated predictor as well as the other clinical predictors.

The result of the model fitting with and without using PV can be found in Table 2. We can immediately see how the significance of the microarray predictor is reduced when 10-fold PV is being used and thus the effect of fitting and testing the model on the same data removed. However, as PV chooses random folds, the results depend on the choice of folds. In order to get a clearer picture of the significance of the microarray predictor, we repeated the 10-fold PV 100 times and averaged the resulting p-values for the analytical and the permutation tests (see Table 3). The analytical test declares the microarray predictor to be significant, however, all 3 permutation test statistics do not give significant results, though the difference of the analytical test to the z-score permutation test is quite small. A possible explanation for these different results is the bias of the analytical test, which we investigate in Sections 4 and 5.2. We also show in Section 6 that the permutation test does not suffer from this problem.

3.2. $L_1$ *penalized logistic regression.* As a second example to illustrate pre-validation, we use a logistic regression model to predict the prognosis of a patient based on the microarray data. Penalized logistic regression (PLR) has been shown to work well for predicting outcomes using gene expression on other datasets [see Zhu and Hastie (2004)] and here we use an $L_1$ penalty as it implicitly performs variable selection on the genes [see Park and Hastie (2007) for an algorithm]. The penalty parameter is chosen such that exactly $5, 10, \ldots, 45$ or 50 genes are included in the model and the optimal number of genes is determined by cross-validation. The exact procedure for pre-validating this cross-validated model is then:

1. Divide the data into $K$ folds.
2. Set aside one fold as the test set, the remaining $K - 1$ folds are the training set.

TABLE 1
*Error rates of the new prediction rules and the clinical predictors. The rates for the new prediction rules are based on* 100 *runs of* 10-*fold CV*

| Predictor | Error rate |
|---|---|
| van't Veer et al. (2002) | 0.321 |
| PLR | 0.379 |
| Grade | 0.333 |
| ER | 0.616 |
| Angio | 0.333 |
| PR | 0.589 |
| Age | 0.654 |
| Size | 0.320 |

TABLE 2
*Summary of the coefficients in the external logistic model with 10-fold PV and without PV using the van't Veer method for prognosis prediction. For each coefficient a test for $\beta = 0$ based on the z-score and the deviance is given. All p-values are for two-sided tests except for the z-score p-value of the van't Veer predictor, which is a one-sided p-value for testing $\beta = 0$ versus $\beta > 0$*

| Predictor  | Method     | Coefficient | SD   | z-score | p-value              | $\Delta$ Deviance | p-value (dev)        |
|------------|------------|-------------|------|---------|----------------------|-------------------|----------------------|
| van't Veer | No PV      | 4.10        | 1.09 | 3.75    | $0.9 \times 10^{-4}$ | 25.01             | $5.6 \times 10^{-7}$ |
|            | 10-fold PV | 1.54        | 0.71 | 2.17    | 0.015                | 5.00              | 0.025                |
| Grade      | No PV      | −0.70       | 1.00 | −0.70   | 0.497                | 0.51              | 0.475                |
|            | 10-fold PV | 0.56        | 0.75 | 0.75    | 0.452                | 0.56              | 0.453                |
| ER         | No PV      | −0.55       | 1.04 | −0.53   | 0.596                | 0.28              | 0.596                |
|            | 10-fold PV | −0.64       | 0.90 | −0.71   | 0.475                | 0.52              | 0.472                |
| Angio      | No PV      | 1.21        | 0.82 | 1.48    | 0.139                | 2.29              | 0.130                |
|            | 10-fold PV | 1.35        | 0.65 | 2.08    | 0.038                | 4.57              | 0.033                |
| PR         | No PV      | 1.21        | 1.06 | 1.15    | 0.251                | 1.39              | 0.238                |
|            | 10-fold PV | 0.43        | 0.83 | 0.51    | 0.609                | 0.27              | 0.606                |
| Age        | No PV      | −1.59       | 0.91 | −1.75   | 0.081                | 3.48              | 0.062                |
|            | 10-fold PV | −1.46       | 0.69 | −2.10   | 0.035                | 4.82              | 0.028                |
| Size       | No PV      | 1.48        | 0.73 | 2.03    | 0.043                | 4.37              | 0.037                |
|            | 10-fold PV | 0.84        | 0.60 | 1.40    | 0.161                | 1.96              | 0.162                |

TABLE 3
*p-values for the van't Veer predictor over 100 runs of the pre-validation procedure. The mean values are reported as well as the percentage below the levels 0.01, 0.05 and 0.1*

| Statistic                 | Mean  | % <0.01 | % <0.05 | % <0.1 |
|---------------------------|-------|---------|---------|--------|
| Analytical z-score        | 0.046 | 15      | 66      | 91     |
| Permutation with $\beta$  | 0.095 | 1       | 27      | 57     |
| Permutation with z-score  | 0.050 | 17      | 62      | 86     |
| Permutation with deviance | 0.139 | 0       | 21      | 42     |

TABLE 4
*Summary of the coefficients in the external logistic model with 10-fold PV and without PV based on the PLR model for prognosis prediction. For each coefficient a test for $\beta = 0$ based on the z-score and the deviance is given. All p-values are for two-sided tests except for the z-score p-value of the PLR predictor, which is a one-sided p-value for testing $\beta = 0$ versus $\beta > 0$*

| Predictor | Method | Coefficient | SD | z-score | p-value | $\Delta$ Deviance | p-value (dev) |
|---|---|---|---|---|---|---|---|
| PLR | No PV | 5.62 | 1.45 | 3.88 | $0.5 \times 10^{-4}$ | 39.33 | $3 \times 10^{-10}$ |
| | 10-fold PV | 0.72 | 0.65 | 1.10 | 0.135 | 1.23 | 0.268 |
| Grade | No PV | 0.69 | 0.99 | 0.70 | 0.487 | 0.48 | 0.488 |
| | 10-fold PV | 0.73 | 0.77 | 0.95 | 0.343 | 0.90 | 0.342 |
| ER | No PV | 0.33 | 1.65 | 0.20 | 0.841 | 0.04 | 0.840 |
| | 10-fold PV | $-0.58$ | 0.87 | $-0.67$ | 0.504 | 0.45 | 0.501 |
| Angio | No PV | 1.05 | 0.91 | 1.15 | 0.250 | 1.34 | 0.247 |
| | 10-fold PV | 1.38 | 0.64 | 2.16 | 0.031 | 4.94 | 0.026 |
| PR | No PV | 1.41 | 1.60 | 0.88 | 0.380 | 0.85 | 0.356 |
| | 10-fold PV | 0.21 | 0.81 | 0.26 | 0.795 | 0.07 | 0.794 |
| Age | No PV | 1.05 | 1.31 | 0.80 | 0.425 | 0.70 | 0.402 |
| | 10-fold PV | $-1.28$ | 0.65 | $-1.97$ | 0.049 | 4.06 | 0.044 |
| Size | No PV | 0.75 | 0.89 | 0.85 | 0.395 | 0.71 | 0.401 |
| | 10-fold PV | 1.13 | 0.58 | 1.93 | 0.053 | 3.83 | 0.050 |

3. Fit the penalized logistic regression model on the training set. Use CV on the training set to find the optimal number of genes.
4. Using the model with the cross-validated number of genes, predict the outcome on the test set.
5. Repeat steps 2–4 for all $K$ folds.
6. Compare the pre-validated predictor to the clinical predictors in a logistic regression model.

As can be seen by this example, PV also works with prediction methods that rely on CV to estimate their parameters. The results using PLR can be seen in Tables 4 and 5. As in the example above, if no PV is being used, the predictor appears to be highly significant. However, using PV, the analytical test gives a one-sided p-value of 0.1349, which is not significant. Using the permutation tests instead confirms this result.

Overall we can see that the method proposed by van't Veer et al. (2002) performs better on this dataset. In the next section we analytically show in a simple case that the analytical test of significance of a pre-validated predictor based on the *t*-statistic is biased.

**4. Analytical results on the bias of tests for pre-validated predictors.** An analytical treatment of the distribution of test statistics in the external model is very difficult in the general case. However, the problem becomes tractable in a simplified setting. Consider PV with $K = n$, that is, leave-one-out PV. Assume that $p < n$ and use a linear regression model for building the new prediction rule. Let there be $e$ other external predictors for the same outcome $y$. Let $X$ be the $n \times p$ matrix with the data used for the new prediction rule.

We assume that $X$ and $y$ have the following distributions:

$$X_{ij} \sim N(0,1) \quad \text{i.i.d. } \forall i = 1, \ldots, n; j = 1, \ldots, p$$

and

$$y_i \sim N(0,1) \quad \text{i.i.d. } \forall i = 1, \ldots, n$$

independent also of $X$. So here our data $X$ is independent of the response $y$ and we can therefore explore the distribution under the null in the external model ($\beta_{PV} = 0$).

4.1. *No other predictors.* For simplicity, let us first consider the case with $e = 0$, that is, no other predictors. As a first step, we need an expression for the prediction using the internal linear model and leave-one-out pre-validation. Here let $H = X(X^T X)^{-1} X^T$ be the projection matrix used in linear regression. Let $D$ be the matrix with the diagonal elements of $H$. Then the leave-one-out pre-validated predictor is

$$\tilde{y} = (I - D)^{-1}(H - D)y =: Py,$$

where $I$ is the identity matrix.

Now use $\tilde{y}$ as the sole predictor in the external model, which is also linear. As there are no other predictors, this may not seem to make much sense, as the hypothesis that there is no relationship between $X$ and $y$ could be tested right away in the internal model. We apply the external model anyway, as it is very instructive as to what the problem is in more complicated settings.

So we now consider the model

$$y = \beta_{PV} \tilde{y} + \varepsilon,$$

where $\varepsilon \sim N(0, \sigma^2 \cdot I)$. Then under these conditions, the following theorem holds:

TABLE 5
*p-values for the PLR predictor over 100 runs of the pre-validation procedure. The mean values are reported as well as the percentage below the levels* 0.01, 0.05 *and* 0.1

| Statistic | Mean | % <0.01 | % <0.05 | % <0.1 |
|---|---|---|---|---|
| Analytical z-score | 0.404 | 0 | 4 | 13 |
| Permutation with $\beta$ | 0.249 | 0 | 3 | 10 |
| Permutation with z-score | 0.275 | 0 | 6 | 14 |
| Permutation with deviance | 0.299 | 1 | 10 | 17 |

THEOREM 1. *Under the assumptions described above, the t-statistic for testing the hypothesis $\beta_{PV} = 0$ has the asymptotic distribution*

$$t = \frac{\hat{\beta_{PV}}}{\hat{sd}(\hat{\beta_{PV}})} \xrightarrow{d} \frac{C-p}{\sqrt{C}} \quad \text{as } n \to \infty,$$

where $C \sim \chi_p^2$.

PROOF. See Appendix A.1. □

As it can be seen here, the statistic is not $t$-distributed as in a regular linear regression. This can lead to biases when the $t$-distribution is used for testing. The size of the bias will be explored numerically later in Section 5.2.

4.2. *Other predictors related to the response $y$.* Now assume that we have several outside predictors for the response. As these are usually based on different data than $X$, we define the distribution of the outside predictors based on $y$ and not on the internal model. So let $Z$ be a $n \times e$ matrix with

$$Z_{ik} = y_i + \gamma_{ik},$$

where $\gamma_{ik} \sim N(0, \sigma_k^2)$ i.i.d. $\forall\, i = 1, \ldots, n;\ k = 1, \ldots, e$. Thus, the additional predictions are perturbed versions of the true response.

The internal model for the prediction of $y$ using $X$ is the same as before. The external linear model now becomes, however,

$$y = \tilde{y}\beta_{PV} + Z\beta + \varepsilon.$$

Again we want to test if $\beta_{PV} = 0$. In a linear model, this is usually done by calculating the t-statistic and calculating the quantile using the $t$-distribution with the right degrees of freedom. The following theorem gives the asymptotic distribution of the $t$-statistic under these assumptions.

THEOREM 2. *Under the setup described above, the t-statistic for testing $\beta_{PV} = 0$ in the external linear model has the asymptotic distribution*

$$t = \frac{\hat{\beta}_{PV}}{\hat{sd}(\hat{\beta}_{PV})} \xrightarrow{d} \frac{(N^T N - p)}{\sqrt{N^T N}} - \frac{N^T A (\mathbf{1}\mathbf{1}^T + \text{Cov}(\gamma))^{-1}\mathbf{1}}{\sqrt{N^T N}(1 - \mathbf{1}^T(\mathbf{1}\mathbf{1}^T + \text{Cov}(\gamma))^{-1}\mathbf{1})}$$
$$\text{as } n \to \infty,$$

where $N \sim N(0, I_p)$, $A = (A_1, \ldots, A_e)$ with $A_k \sim N(0, \sigma_k^2 \cdot I_p))$, $\mathbf{1} = (1, \ldots, 1)^T \in \mathbb{R}^e$ and $\text{Cov}(\gamma) = \text{diag}(\sigma_1^2, \ldots, \sigma_e^2)$.

PROOF. See Appendix A.2. □

We can see that the asymptotic distribution of the $t$-statistic is not a $t$ or normal distribution, as we already observed in the simple case above without external predictors.

In the next section, by using simulations, we will investigate the extent of the bias when the testing is done using a $t$-distribution.

## 5. Models, bias and permutation test.

5.1. *Models used in the simulations.* In the section above we have seen that in the simple case where the internal and external models are linear regressions, the $t$-statistic does not have its usual distribution. We expect that the same is true for more complicated scenarios, which are not tractable analytically. In order to investigate the amount of bias in more complex settings, we used the following 3 model combinations in our simulations.

5.1.1. *Linear–linear.* This is the most simple model and was also used in the analytical analysis. Here, the internal and external models are standard linear regressions. Let $n$ be the number of subjects and $p$ be the number of predictors for the internal model. Let $e$ be the number of external predictors. Then the internal predictors are a matrix $X$ which is generated as

$$X_{ij} \sim N(0,1) \qquad \text{i.i.d. } i=1,\ldots,n, j=1,\ldots,p.$$

With $\beta \in \mathbb{R}^p$ a user supplied vector, the response is generated as

$$y \sim N(X\beta, I \cdot \sigma_I^2).$$

From this true response, the external predictors are derived as

$$Z_{ik} \sim N(y_i, \sigma_E^2) \qquad \text{i.i.d. } i=1,\ldots,n, k=1,\ldots,e.$$

The rationale for simulating the external predictors as a perturbation of the truth rather than the underlying model is that the external predictors would be derived using different models and may be targeting other aspects of the phenomenon such that the underlying model here would not apply to them. From this perspective, modeling them as a noisy version of the truth seems more appropriate. For simplicity, we always choose $\sigma_I^2 = \sigma_E^2 = 1$ in the simulations.

5.1.2. *Lasso–linear.* This model is an extension of the previous one. The predictor matrix $X$ is generated in exactly the same way as before. However, only the first $s$ components of $\beta$ are being supplied by the user. The other $p-s$ components are set to 0 to ensure sparseness. The external predictors are then generated from $y$ as described above.

For analyzing this artificial data, an internal lasso regression model will be used. The external model is linear regression as before. The internal model will be fit using the LARS algorithm [see Efron et al. (2004)], ensuring that the fitted model contains exactly a prespecified number $l$ of nonzero coefficients. $l$ is chosen by the user. More sophisticated methods are possible, but outside the scope of this paper.

5.1.3. *Linear Discriminant Analysis (LDA)–Logistic.* This model is intended to simulate something close to realistic applications on microarray data. Again, there are $n$ observations, which are divided into 2 groups with $n_1$ and $n_2$ members ($n_1 + n_2 = n$). Also, $p$ predictors (genes) will be generated for each observation independently. However, for the first $s$ out of the $p$ genes, the means will be different. For the first group, $\mu_{ij} = 0 \, \forall i, j$, where $i$ refers to the observation and $j$ to the genes. For the second group of $n_2$ observations, the first $s$ genes will have $\mu_{ij} = \mu > 0$, a positive offset in the mean from the same genes in the first group. All others genes will also have mean 0 in the second group as well. Then we simulate the microarray data as

$$X_{ij} \sim N(\mu_{ij}, \sigma^2).$$

The external predictors are then generated by switching the label of the $y_i$ independently with probability $p_E$.

In the internal model, first a number $g$ of predictors is selected by choosing the predictor with the largest correlation with the response. Then an LDA model is fit to the chosen $g$ predictors. The number $g$ will be supplied by the user. As above, automatic choices are possible, but as we just want to demonstrate the performance of PV, we keep $g$ fixed. In the external model, standard logistic regression is used. For simplicity, we again choose $\sigma = 1$.

5.2. *Simulation of the type I error under the null.* In each of the scenarios described above, we simulate artificial data and perform the PV algorithm 100,000 times (without the permutation test). The analytical p-value of the pre-validated predictor is used to decide if the null hypothesis is rejected ($t$-statistic in linear regression model, z-score in logistic regression). Based on the simulations, the type I error of the analytical test is estimated (see Table 6).

The analytical tests in the external models show substantial upward and downward bias in the tested scenarios, depending on the choice of parameters. For the type I error level 0.01, this upward bias can double the size of the test and it is also substantial at level 0.05.

The remedy for this problem is a permutation test.

5.3. *The permutation test.* As we have just seen, the standard analytical test in the external models used (here linear and logistic) do not achieve their nominal level when they are being applied to pre-validated predictors. This can have serious consequences on the outcome of the test. A permutation test is a procedure that is very robust with respect to this problem.

TABLE 6
*Type* I *error in various scenarios. Each estimate is based on 100,000 simulations, giving an SD of* ≤0.005. *The most extreme values for each scenario are in bold*

| Scenario | Parameters | CV-folds | Type I error | | |
|---|---|---|---|---|---|
| | | | $\alpha = 0.01$ | $\alpha = 0.05$ | $\alpha = 0.1$ |
| Linear–Linear | $n=10, p=5, k=1, \beta=0$ | 5 | 0.022 | 0.079 | 0.137 |
| | | 10 | **0.024** | 0.080 | 0.139 |
| | | $n$ | 0.023 | **0.083** | **0.140** |
| | $n=20, p=5, k=1, \beta=0$ | 5 | 0.018 | 0.069 | 0.123 |
| | | 10 | 0.017 | 0.066 | 0.120 |
| | | $n$ | 0.018 | 0.067 | 0.119 |
| | $n=50, p=5, k=1, \beta=0$ | 5 | 0.016 | 0.064 | 0.115 |
| | | 10 | 0.016 | 0.062 | 0.111 |
| | | $n$ | 0.015 | 0.060 | 0.109 |
| Lasso–Linear | $n=10, p=100, k=1, \beta=0, s=0, l=5$ | 5 | **0.008** | **0.033** | **0.062** |
| | | 10 | 0.011 | 0.040 | 0.072 |
| | $n=10, p=100, k=1, \beta=0, s=0, l=10$ | 5 | 0.010 | 0.040 | 0.074 |
| | | 10 | 0.016 | 0.053 | 0.091 |
| | $n=30, p=100, k=1, \beta=0, s=0, l=5$ | 5 | 0.012 | 0.040 | 0.071 |
| | | 10 | 0.014 | 0.046 | 0.076 |
| | $n=30, p=100, k=1, \beta=0, s=0, l=10$ | 5 | 0.016 | 0.054 | 0.092 |
| | | 10 | 0.021 | 0.065 | 0.105 |
| | $n=30, p=100, k=1, \beta=0, s=0, l=20$ | 5 | 0.020 | 0.065 | 0.112 |
| | | 10 | **0.030** | **0.081** | **0.128** |
| LDA–Logistic | $n=20, p=1000, k=1, \beta=0, s=0, g=10$ | 5 | **0.003** | **0.025** | **0.076** |
| | | 10 | 0.0096 | 0.047 | 0.100 |
| | $n=40, p=1000, k=1, \beta=0, s=0, g=10$ | 5 | 0.018 | 0.072 | 0.122 |
| | | 10 | 0.036 | 0.106 | 0.158 |
| | $n=80, p=1000, k=1, \beta=0, s=0, g=10$ | 5 | 0.019 | 0.071 | 0.122 |
| | | 10 | **0.053** | **0.126** | **0.179** |

The external predictors have usually been used and validated in this context before, so we were not concerned with evaluating their performance. In any case, extending the permutation test to cover them as well is straightforward. The variables that we have as input is the response $y$, the internal predictors $X$ and the external predictors $Z$. As there is a relationship between $y$ and $Z$, we do not permute $y$ but instead the rows of $X$. Then, the pre-validation procedure is used and a test statistic in the external model collected (say $\beta$ or $t$). This permutation is repeated often enough to get a sufficiently large sample of the test statistic (here usually 500 or 1000 permutations). The p-value is then estimated as the fraction of the permutation test statistic larger or equal to the observed test statistic (no randomiza-tion on the boundary). As the pre-validated predictor is a prediction for the response $y$, we expect its coefficient to be positive and therefore use a one-sided p-value (as we already did for the analytical test).

The external predictors $Z$ remain unchanged by the permutation, even if they were derived with or are otherwise dependent on the internal data $X$. Another possibility would be to model the dependency of $Z$ on $X$ and also change $Z$ when $X$ is being permuted. We chose not to use this approach for the following reasons:

- The model for deriving $Z$ from $X$ may be unknown. This would be the case when the researcher was just provided with the clinically relevant information.
- The exact underlying relationship between $Z$ and $X$ may be unknown. If, say, $X$ is microarray data and $Z$ is derived from nonmicroarray data (e.g., blood samples, tumor measurements, ...), it is still likely that there is some relationship between these data types. This relationship may be unknown so that it would be impossible to assess the effect of permuting $X$ on $Z$.

These problems make our method much easier to implement in practice. Furthermore, the simulation results show that the method works well even for dependent $Z$ and $X$.

**6. Simulation results.** In this section we explore whether the permutation test achieves the intended level and what effect it has on the power of the test compared to the analytical solution. For this, artificial datasets according to the 3 scenarios described above are created and analyzed.

6.1. *Level of the permutation test.* For estimating the level of the test under the null hypothesis, the internal predictors $X$ will be independent of the response and the external predictors $Z$. Several different parameter combinations will be used for this task. For each scenario and parameter choice, 1000 simulations were used where each test was based on 500 permutations.

All estimates are well within 2 standard deviations of their target value, so we see that the permutation tests are unbiased. The simulated levels of the permutation tests can be found in Tables 1, 2 and 3 of the Supporting Online Material (SOM) [Höfling and Tibshirani (2008)]. The standard error for the $\alpha = 0.01$ estimate is 0.003, for $\alpha = 0.05$ it is 0.007 and for $\alpha = 0.1$ the standard error is 0.009.

6.2. *Power.* The same scenarios that were used for estimating the level of the permutation tests will also be used to estimate the power under the alternative. As there is no distinct alternative hypothesis, several different choices will be used, depending on the specific scenario.

One of the most interesting aspects of this simulation is to compare the power of the permutation test to the power of the standard analytical test. However, as the analytical test is biased (usually upward), a straightforward comparison using the nominal test levels is inappropriate. In order to adjust for the bias, the simulations in the same scenario and parameters under the null hypothesis will be used. For each nominal level, a new cutoff for the p-values will be estimated such that the level of the analytical test is equal to its nominal level. This cutoff will then also be used to estimate its power.

The power of the permutation test is in most cases very close to the power of the analytical test and sometimes even higher (although this may be a random occurrence). So, there does not seem to be a serious problem with loss of power when comparing the permutation tests to the analytical test. The simulated results can be seen in Tables 4, 5 and 6 of the SOM [Höfling and Tibshirani (2008)]. As before, the estimates are based on 1000 simulations, each of which used 500 permutations for the tests. Here, the maximum standard deviation for the test is achieved for a power of 0.5, in which case the SD is 0.016.

However, the picture as to which choice of test statistic and number of folds to use for the permutation test is not very clear. For the Linear–Linear model, we used 5-fold PV, 10-fold PV, leave-on-out PV and permutation tests without PV ($K=1$). For the other model, due to computation time constraints, we only used 5- and 10-fold PV as well as no PV. In the Linear–Linear scenario, leave-one-out PV performs slightly better than 5-fold and 10-fold PV. However, in all but the simplest models, performing leave-one-out PV comes with a serious increase in computation time so that just using 5- or 10-fold PV may be considered appropriate.

In some instances, the permutation test using no PV showed a lot more power than 5- or 10-fold PV permutation tests. However, especially in the LDA–Logistic model, the test without PV had power even below the nominal level of the test. This can be explained by overfitting the data, leading to perfect separation of the classes even if there is no relationship between the class labels $y$ and the internal predictors $X$. In these cases, the permutation test without PV does not give useable results.

Therefore, using the 5-fold (or 10-fold) PV permutation test is the most reliable procedure, achieving the nominal level of the test without compromising power with respect to the analytical test. The choice of test statistic depends on the specific application, but all standard statistics we used had acceptable performance.

6.3. *Performance of the estimator for the pre-validated coefficient.* When the new prediction rule turns out to be a significant improvement over the performance of the old prediction rules, the value of the coefficient of the new predictor compared to the coefficients of the old predictors indicates how well the new predictor performs. Therefore, it is important to know how well PV estimates the coefficient of the new prediction rule.

In order to have a comparison that is fair and relevant with respect to the amount of data available, we estimate the coefficient using PV over 1000 simulation runs in the scenarios presented above. As a benchmark method, we treat the dataset the PV was performed on as a training set to estimate the new prediction rule and do the comparison to the other prediction rules on an independently simulated test dataset of the same size as the training data. Our primary concern is that the coefficient estimated using PV is roughly unbiased w.r.t. the benchmark. The most straightforward approach would be to compare the mean over the simulations of the estimated coefficient using PV and using the benchmark. However, in the LDA–Logistic scenario, occasionally perfect separation occurs which makes the estimated coefficients extremely large. Mean-unbiasedness is not applicable in this case and we decided to use median-unbiasedness instead. As the difference between mean and median is quite small in all other scenarios and the median is more robust, we used the median in the remaining scenarios as well [for results see Table 10 of SOM; Höfling and Tibshirani (2008)].

In general, PV tends to underestimate the coefficient compared to the benchmark. The size of the underestimation depends on the scenario and the number of folds used in PV. The performance in the Linear–Linear model is very good with hardly any bias at all. For the Lasso–Linear and the LDA–Logistic scenario, the bias is bigger. The difference of the estimates for 5-fold and 10-fold PV show that at least part of the bias is due to the smaller training set used for deriving the prediction rule in PV. The bias also decreases with increasing number of observations, which can also be explained this way, as removing a certain percentage of observations has a smaller perturbing effect on the prediction rule when the total number of observations is large. Overall, PV does a good job of estimating the coefficient of the new prediction rule.

**7. Discussion.** The problem often arises that, with a limited amount of data, one wants to find a prediction rule and verify its usefulness on the same dataset. Often, splitting the data into separate training and test sets [as in, e.g., Chang et al. (2005)] is not feasible as there may not be enough samples to achieve acceptable prediction performance and have enough observations left to compare additional clinical predictors to the new prediction rule. Pre-validation is a useful method to fill this gap and evaluate the significance and prediction performance of the newly developed prediction rule. However, we have found that using the standard analytical tests with the pre-validated predictor can yield a test with level above the nominal level.

The permutation test approach to the pre-validated predictor addresses the bias problem of the analytical test without compromising power and is therefore a more reliable way for assessing whether the new prediction rule is an improvement over previously established predictors. Its main drawback is that it is very computer-intensive, requiring us to refit the pre-validation model for every permutation. This can be a problem for especially large datasets. However, this will not often be a significant problem and the simple structure of the algorithm makes it easily accessible to parallelization to reduce computation time.

It might be possible to develop an analytical test that accounts for the special structure of the pre-validated predictor. However, it is unclear if an analytical solution exists that holds for a large number of models. Since the internal models are usually tailored to the specific problem at hand, having to derive analytical solutions on a case by case basis would be very difficult. We believe that the permutation test is the best method currently available for the problem.

## APPENDIX: PROOFS

**A.1. Case of no outside predictors.** For the proof, we first need a lemma:

LEMMA A1. *Let $X_{ij}$ be i.i.d. $N(0,1)$ for $i = 1, \ldots, n$ and $j = 1, \ldots, p$. Let $H = \mathrm{Proj}(X) = X(X^T X)^{-1} X^T$ and $D = \mathrm{diag}(H)$. Then $d_{ii} \sim O_P(n^{-1})$.*

PROOF. By the strong law of large numbers, $\frac{1}{n} X^T X \to I_p$ a.s. and as taking the inverse of a matrix is a continuous operation,

$$n(X^T X)^{-1} \to I_p \quad \text{a.s.}$$

Therefore,

$$nd_{ii} = nx_i(X^T X)^{-1} x_i^T \xrightarrow{d} \chi_p^2$$

by continuous mapping, where $x_i$ is the $i$th row of $X$. □

Also note that as $\mathrm{trace}(H) = \sum_i d_{ii} = p$, we have that $\mathrm{Cov}(d_{ii}, d_{jj}) < 0 \ \forall i \neq j$.

Now let us move on to the proof of Theorem 1.

PROOF OF THEOREM 1. Let the SVD of $X$ be

$$X = UEV^T,$$

with $U \in \mathbb{R}^{n \times p}$ orthogonal, $E \in \mathbb{R}^{p \times p}$ diagonal and $V \in \mathbb{R}^{p \times p}$ orthogonal. Then we can write $H = UU^T$, therefore, the leave-one-out pre-validated predictor is

$$\tilde{y} = (I - D)^{-1}(UU^T - D)y \quad \text{and} \quad \hat{\beta}_{PV} = \frac{\tilde{y}^T y}{\tilde{y}^T \tilde{y}}.$$

Evaluating the numerator, we get

$$\tilde{y}^T y = y^T (UU^T - D)(I - D)^{-1} y$$
$$= y^T UU^T y + y^T UU^T ((I - D)^{-1} - I) y$$
$$- y^T D((I - D)^{-1} - I) y - y^T D y$$
$$\xrightarrow{d} N^T N + 0 - p \quad \text{as } n \to \infty,$$

where $N \sim N(0, I_p)$. This holds as $U^T y \sim N(0, I_p)$. The second term converges to 0 as $((I - D)^{-1} - I) \sim O_P(n^{-1})$ and $U^T y = N$ is bounded in probability. The third term converges to 0 in probability as $D((I - D)^{-1} - I) \sim O_P(n^{-2})$. For the fourth term observe that $E(y^T D y) = E(E(y^T D y | X)) = E(\sum d_{ii}) = p$. As $\text{Cov}(d_{ii}, d_{jj}) < 0$ for $i \neq j$, it is easy to show that $y^T D y \xrightarrow{P} p$.

For the denominator, we get

$$\tilde{y}^T \tilde{y} = y^T (UU^T - D)(I - D)^{-2} (UU^T - D) y$$
$$= N^T N + N^T U^T ((I - D)^{-2} - I) U N$$
$$- 2 y^T D (I - D)^{-2} U N + y^T D^2 (I - D)^{-2} y.$$

Here, the first term is $N^T N$ as above and the other terms converge to 0. The second and third summand converge to 0 as $(I - d)^{-2} - I \sim O_P(n^{-1})$ and $D(I - D)^{-2} \sim O_P(n^{-1})$ and for the fourth term we use that $D^2 (I - D)^{-2} \sim O_P(n^{-2})$.

Now that we have the distribution of the numerator and denominator of $\hat{\beta}_{PV}$, consider $\hat{sd}(\hat{\beta}_{PV})$. This is estimated as

$$\hat{sd}(\hat{\beta}_{PV} 0) = \hat{\sigma} \sqrt{\tilde{y}^T \tilde{y}}.$$

Only $\hat{\sigma}$ is left to treat, for which we can write

$$\hat{\sigma}^2 = \frac{1}{n - 1} (y - \hat{\beta}_{PV} \tilde{y})^T (y - \hat{\beta}_{PV} \tilde{y})$$
$$= \frac{1}{n - 1} (y^T y - 2 \hat{\beta}_{PV} \tilde{Y}^t y + \hat{\beta}_{PV}^2 \tilde{y}^T \tilde{y}).$$

We know that $\frac{1}{n-1} y^T y \to 1$ a.s. The other terms go to 0, as it has been shown above that the second and third summand inside the bracket is bounded in probability.

So putting all this together yields the desired result. $\square$

**A.2. Case with outside predictors.** The proof of Theorem 2 is along the lines of the proof for Theorem 1, but with more complicated algebra.

First recall a well-known fact about the inverse of matrices. Assume we have a matrix with blocks of the form

$$M = \begin{pmatrix} A & B \\ C & D \end{pmatrix},$$

where $A$ and $D$ are nonsingular square-matrices. Then we can write the inverse $M^{-1}$ as

$$M^{-1} = \begin{pmatrix} (A - BD^{-1}C)^{-1} & -(A - BD^{-1}C)^{-1}BD^{-1} \\ -D^{-1}C(A - BD^{-1}C)^{-1} & D^{-1} + D^{-1}C(A - BD^{-1}C)^{-1}BD^{-1} \end{pmatrix}.$$

The proof of Theorem 2 is then:

PROOF OF THEOREM 2. Let $\beta = (\beta_{PV}, \beta_1^T)^T$ and $W = (\tilde{y}, Z)$. Then

$$\hat{\beta} = (W^T W)^{-1} W^T y \quad \text{where } W^T W = \begin{pmatrix} \tilde{y}^T \tilde{y} & \tilde{y}^T Z \\ Z^T \tilde{y} & Z^T Z \end{pmatrix}$$

and as we are only interested in $\hat{\beta}_{PV}$, this can be written as

$$\hat{\beta}_{PV} = (\tilde{y}^T \tilde{y} - \tilde{y}^T Z (Z^T Z)^{-1} Z^T \tilde{y})^{-1} (\tilde{y}^T y - \tilde{y}^T Z (Z^T Z)^{-1} Z^T y),$$

using the formula for inverses of block matrices. Also define $\mathbf{1} = (1, \ldots, 1)^T \in \mathbb{R}^e$. Then

$$\frac{1}{n} Z^T Z = \frac{1}{n} (y \cdot \mathbf{1}^T + \Gamma)^T (y \cdot \mathbf{1}^T + \Gamma)$$

$$= \frac{1}{n} (y^T y \mathbf{1} \mathbf{1}^T + 2 \cdot \mathbf{1} y^T \Gamma + \Gamma^T \Gamma)$$

$$\xrightarrow{P} \mathbf{1}\mathbf{1}^T + 0 + \text{Cov}(\gamma),$$

where $\Gamma_{ik} = \gamma_{ik}$ is the matrix of random errors of the external predictors and the convergence follows by the weak law of large numbers.

Also,

$$\frac{1}{n} Z^T y = \frac{1}{n} (\mathbf{1} y^T y + \Gamma^T y) \xrightarrow{P} \mathbf{1} + 0,$$

again using the weak law of large numbers and the independence of $\Gamma$ and $y$. Furthermore $Z^T \tilde{y} = \mathbf{1} y^T \tilde{y} + \Gamma^T \tilde{y}$. As we already know that $y^T \tilde{y} \xrightarrow{d} N^T N - p$ where $N \sim N(0, I_p)$, we only have to determine the distribution of

$$\Gamma^T \tilde{y} = \Gamma^T (I - D)^{-1} (H - D) y$$

$$= \Gamma^T (I - D)^{-1} U U^T y - \Gamma^T (I - D)^{-1} D y$$

$$\xrightarrow{d} A^T N - 0,$$

where $N = U^T y \sim N(0, I_p)$ and $U^T (I - D)^{-1} \Gamma \xrightarrow{d} A = (A_1, \ldots, A_e)$ with $A_k \sim N(0, \sigma_k^2 \cdot I_p))$ i.i.d. So $Z^T \tilde{y}$ converges in distribution to

$$Z^T \tilde{y} \xrightarrow{d} N^T N - p + A^T N.$$

So combining the previous results, we have

$$\tilde{y}^T Z (Z^T Z)^{-1} Z^T \tilde{y} = \frac{1}{n} \left( \tilde{y}^T Z \left( \frac{1}{n} Z^T Z \right)^{-1} Z^T \tilde{y} \right) \xrightarrow{P} 0,$$

as the term inside the brackets is bounded in probability. Also,

$$\tilde{y}^T Z(Z^T Z)^{-1} Z^T y$$
$$= \tilde{y}^T Z\left(\frac{1}{n} Z^T Z\right)^{-1} \frac{1}{n} Z^T y$$
$$\xrightarrow{d} (\mathbf{1}^T(N^T N - p) + N^T A)(\mathbf{11}^T + \text{Cov}(\gamma))^{-1}\mathbf{1}.$$

Combining all this, we have that

$$\hat{\beta}_{PV} \xrightarrow{d} \frac{N^T N - p - (\mathbf{1}^T(N^T N - p) + N^T A)(\mathbf{11}^T + \text{Cov}(\gamma))^{-1}\mathbf{1}}{N^T N}$$
$$= \frac{(N^T N - p)(1 - \mathbf{1}^T(\mathbf{11}^T + \text{Cov}(\gamma))^{-1}\mathbf{1}) - N^T A(\mathbf{11}^T + \text{Cov}(\gamma))^{-1}\mathbf{1}}{N^T N}.$$

In order to get the distribution of the $t$-statistic, the distribution of

$$\hat{sd}(\hat{\beta}_{PV}) = \sqrt{(W^T W)_{11}^{-1}} \hat{\sigma}$$

is needed. First, consider $(W^T W)_{11}^{-1}$:

$$(W^T W)_{11}^{-1} = (\tilde{y}^T \tilde{y} - \tilde{y}^T Z(Z^T Z)^{-1} Z^T \tilde{y})^{-1} \xrightarrow{d} (N^T N)^{-1}$$

as $\tilde{y}^T Z(Z^T Z)^{-1} Z^T \tilde{y} = \frac{1}{n}\tilde{y}^T Z(\frac{1}{n} Z^T Z)^{-1} Z^T \tilde{y} \xrightarrow{P} 0$. Next determine the asymptotic distribution of $\hat{\sigma}$:

$$\hat{\sigma} = \frac{1}{n - e - 1}(y - \hat{y})^T(y - \hat{y})$$
$$= \frac{1}{n - e - 1}(y^T y - y^T W(W^T W)^{-1} W^T y).$$

As before, $\frac{1}{n-e-1} y^T y \xrightarrow{P} 1$. For the second term, first observe that

$$\frac{1}{n} W^T y \xrightarrow{P} \begin{pmatrix} 0 \\ \mathbf{1} \end{pmatrix}.$$

For $\frac{1}{n}(W^T W)^{-1}$, it is simple to show that all elements are asymptotically bounded in probability. For $\hat{\sigma}$, only the bottom right block is needed, where

$$\frac{1}{n}(W^T W)_{22}^{-1} \xrightarrow{P} (\mathbf{11}^T + \text{Cov}(\gamma))^{-1} \quad \text{as } n \to \infty.$$

Therefore,

$$\hat{\sigma} \xrightarrow{d} 1 - \mathbf{1}^T(\mathbf{11}^T + \text{Cov}(\gamma))^{-1}\mathbf{1}.$$

Combining these results yields the claim. □

**Acknowledgments.** We thank anonymous referees and the editor for helpful comments and suggestions.

## SUPPLEMENTARY MATERIAL

**Supporting online material for "A study of pre-validation"** (doi: [10.1214/08-AOAS152SUPP](10.1214/08-AOAS152SUPP); .pdf).

Department of Statistics  
Stanford University  
Stanford, California 94305  
USA  
E-mail: [hhoeflin@stanford.edu](hhoeflin@stanford.edu)

Departments of Health Research & Policy, and Statistics  
Stanford University  
Stanford, California 94305  
USA  
E-mail: [tibs@stat.stanford.edu](tibs@stat.stanford.edu)